\documentstyle[preprint,aps]{revtex}

\let\s=\sigma

\let\S=\Sigma

\newcommand{\be}{\begin{equation}}
\newcommand{\ee}{\end{equation}}
\newcommand{\bea}{\begin{eqnarray}}
\newcommand{\eea}{\end{eqnarray}}

\newcommand{\del}{\partial}
\newcommand{\td}{$\tilde{D}0$-branes\ }

\newcommand{\nbox}{{\,\lower0.9pt\vbox{\hrule \hbox{\vrule height 0.2 cm \hskip
0.2 cm \vrule height 0.2 cm}\hrule}\,}}

\begin{document}
\begin{titlepage}
\begin{center}
\vskip .2in
\hfill
\vbox{
    \halign{#\hfil         \cr
           hep-th/9704123 \cr
           UTTG-14-97      \cr
           SLAC-PUB-7462\cr
           April 1997    \cr
           }  
      }   
\vskip 0.5cm
{\large \bf M(atrix) String Theory on K3}\\
\vskip .2in
{\bf Willy Fischler$^{\diamondsuit}$} \footnote {e-mail address:
fischler@utaphy.ph.utexas.edu},
{\bf Arvind Rajaraman$^{\clubsuit}$} \footnote{e-mail address:
arvindra@dormouse.stanford.edu}\footnote{Supported in part by the Department of
Energy
under contract no. DE-AC03-76SF00515.}\\
\vskip .25in
{\em
$^{\diamondsuit}$Theory Group,Department of Physics, University of Texas,
Austin,TX 78712.\\
$^{\clubsuit}$Stanford Linear Accelerator Center,
     Stanford, CA 94309 \\}
\vskip 1cm
\end{center}
\begin{abstract}
We conjecture that M-theory compactified on an ALE space (or K3)
 is described  by 0-branes moving on the ALE space. We give evidence for
this by showing that if we compactify another circle, we recover string theory
on the ALE space. This guarantees that in the large N limit, the
matrix model correctly describes the force law between gravitons moving in
an ALE background.
We also show the appearance in M(atrix) theory of the duality of M-theory
on K3 with the heterotic string on a three-torus.
\vskip 0.5cm
\begin{center}
Submitted to Nuclear Physics B.
\end{center}
 \end{abstract}
\end{titlepage}
\newpage

\newpage
\bigskip

\section{Introduction}
M(atrix) theory \cite{bfss} has been proposed as a nonperturbative formulation
of M-theory.
The conjecture is that M-theory in the infinite momentum frame is
described by a Hamiltonian that consists of  zero-branes and their
interactions. These
zero-branes are the carriers of longitudinal momentum. We shall refer
to these partons as \td . This conjecture has passed many tests. The
Hamiltonian contains
11-dimensional supergravitons and supermembranes in its spectrum and reproduces
their interactions \cite{bfss,DKPS,lifsch,polchpoul}. It also possesses
T-duality upon compactification \cite{tdual,grt}.
In addition, upon compactifying on a circle and shrinking the circle to
zero radius, M(atrix) theory has been shown \cite{Motl,IIAbs,IIAVV,banksmotl}
to contain multistring states. These strings interact according to the usual
light-cone
interactions and the proper scaling between the coupling constant
and  the radius is recovered \cite{IIAVV}.

It is important to consider compactifying the M(atrix) model on more
complicated
surfaces which preserve less supersymmetry. The natural first step is
to consider the simplest Calabi-Yau
space, the four-dimensional $K3$ surface. It is a reasonable
guess that M-theory on $K3$
in the infinite momentum frame is described
by the dynamics of \td \ moving on $K3$ \cite{enhance}.

However, in \cite{ods}, the authors computed the force law between two  \td \
moving
on $K3$. They found that the $v^4$ part of the force was not the same as the
force expected from supergraviton exchange. This appeared to signal a potential
problem with this model.

We believe that this disagreement is not a fatal one. There is no
nonrenormalization theorem for the gravitational coupling constant
in ${\cal N}=2$ theories. Therefore the coupling
constant could easily be different at short and large distance
scales.
In any case, for the connection to M(atrix) theory, we need to consider
 the interaction of   bound states of large numbers of \td since  it is only
in the large $N$ limit that one recovers the force
law derived from supergravity.

In order to show how the large $N$ limit is involved in this context and to
simplify the discusssion, we will begin by following the
proposal of \cite{enhance} for a matrix model of
M-theory on an ALE surface. We will then compactify an additional circle. In
the limit where this circle
is small, (keeping the ALE size fixed in string units), we will show how to
recover type IIA strings propagating
in an ALE background.   String theory
then guarantees that the gravitational force law is exactly reproduced in
M(atrix) theory without the need for additional degrees of freedom beyond the
\td . Since the force law between gravitons is reproduced in
light-cone string theory only when $N\rightarrow \infty$, there is
no conflict with the results of \cite{ods}.

Finally, we subject this model to another test . In the
limit when the volume of the $K3$ shrinks to zero, string duality
predicts that we should
get the dynamics of heterotic strings on $T^3$. We show that this
is indeed the case, thus providing another consistency
check of this model.

 \section{M(atrix) String Theory}
We will review the calculation of \cite{IIAVV}, expressing it in terms that
will be suitable for
us.

The bosonic part of the M(atrix) theory hamiltonian is
\be
H=R_{11} \ tr\ (\Pi_i^2+[X^i,X^j]^2)
\ee

 On compactification on a circle of radius $R_9$, the M(atrix) model  is
described by a
1+1 dimensional gauge theory \cite{wati} with the Hamiltonian
\be
H=\int^{2\pi}{d\s\over R_9} \  tr\  (\Pi_i^2+R_9^2\ (DX^i)^2+{1\over
R_9^2}E^2+[X^i,X^j]^2)
\ee

Rescaling $X^i\rightarrow R_9^{-1/2}X^i$, we get
\be
H=\int^{2\pi}{d\s} \  tr\  \bigg(\Pi_i^2+(DX^i)^2+{1\over
R_9^3}(E^2+[X^i,X^j]^2)\bigg)
\ee

For small $R_9$, the type IIA limit, we find that the terms $[X,X]$ and $E$
must be
set equal to zero. The condition $[X,X]=0$ implies that we are on the moduli
space of the theory where the $X^i$ commute. The condition $E=0$ implies that
the gauge dynamics is
such that only gauge-singlet states survive. The objects which are
gauge-invariant
are the eigenvalues of the matrices $X^i$. Accordingly, the small $R_9$
dynamics
is reduced to the dynamics of the eigenvalues of $X^i$. This can be described
as a theory with $Z_N$ gauge invariance.

The $X^i$ are functions of the coordinate $\s$ in the YM theory. Due to
the residual $Z_N$ symmetry, they do not have to be periodically identified,
but rather only identified upto a permutation of the eigenvalues. This
effectively multiplies the length of the string by $N$. We can make strings
of any longitudinal momentum by this procedure.

The world sheet Hamiltonian reduces to
\be
H=\int^{2\pi}{d\s}\  tr\  (\Pi_i^2+(\del X^i)^2)
\ee
Thus the long strings which are
produced have the dynamics of $IIA$ strings.

We now turn to the problem we are interested in, which is
the M(atrix) model compactified on $K3$. We shall start with a simpler
case, where we have an ALE surface instead of $K3$. The conjecture is
then that M-theory on an ALE surface is described by the dynamics
of 0-branes on the ALE. We will start by working in the
orbifold limit of the ALE surface. Coordinates will be chosen so that
$(x_6,x_7,x_8,x_9)$ are the directions along the ALE surface.

The dynamics of 0-branes on an ALE space was studied in great detail in
\cite{dm,jm}. For illustrative purposes, we will take the example $R^4/Z_2$.
In this case,
each 0-brane has an image.
We therefore take $N$ 0-branes and their images under $Z_2$ moving on $R^4$.
One
then quantizes the open strings connecting these 0-branes and then imposes
invariance under $Z_2$.

There are two types of strings connecting these branes, those with
polarizations
along the ALE surface (which we will call $\psi^a,a=6,7,8,9$) and those with
polarizations
perpendicular to the ALE surface (which we call $X^i,i=0,1,2,3,4,5$).
Each of them is represented by a matrix in $U(2N)$.
One then imposes the $Z_2$ projection on these states.

We can schematically write the $Z_2$ projection as
\be
U X^i U^{-1}=X^i\qquad U \psi^a U^{-1}=-\psi^a
\ee
where $U$ is a unitary operator representing the $Z_2$.

We can represent the solution of these equations by
\be
X^i=\left(\matrix{
A^i & B^i \cr
B^i & A^i \cr
}\right)\qquad
\psi^a=\left(\matrix{
C^a & D^a \cr
-D^a & -C^a \cr
}\right)
\ee
where $A^i,B^i,C^a,D^a$ are $N\times N$ matrices.

In order to derive the Hamiltonian describing the dynamics of zero-branes, it
is
useful to start with a 6-dimensional theory since the zero-branes
move on $R^6\times$ ALE.
The resulting theory, thought of as a ${\cal N}=1,d=6$ theory,
has $U(N) \times U(N)$ gauge symmetry, and there are (in addition to the
vector multiplet) , two hypermultiplets transforming in the $(N,\bar{N})$
representation.
The theory on the world-volume of the zero-branes is then the dimensional
reduction of this
gauge theory to 0+1 dimensions.

We are interested in the case where in addition, we compactify on a circle
of radius $R_9$.  We can T-dualize along
the circle in order to get a theory of 1-branes wrapped
on the dual circle. After doing this the Hamiltonian is \cite{dm,jm}
\bea
H=\int{d\s\over R_9}\ tr\ \Bigl(R_9^{-2}E^2
+P_X^2+P_\psi^2+&&R_9^2(DX^i)^2+R_9^2(D\psi^a)^2+\nonumber\\
\sum_{a,b}&&{|[\psi^a,\psi^b]|^2}+
2\sum_{a,i}{|[\psi^a,X^i]|^2}+\sum_{i,j}{|[X^i,X^j]|^2}\Bigr)
\eea
where commutators are taken in the $U(2N)$ matrices.

We rescale $X^i\rightarrow R_9^{-1/2}X^i,\psi^a\rightarrow
R_9^{-1/2}\psi^a$.The
Hamiltonian is then
\bea
H=\int{d\s}\ tr\ \Bigl(P_X^2+P_\psi^2&&+(DX^i)^2+(D\psi^a)^2+{1\over
R_9^3}E^2\nonumber\\
{1\over
R_9^3}&&\Bigl[{\sum_{a,b}{Tr|[\psi^a,\psi^b]|^2}+
2\sum_{a,i}{Tr|[\psi^a,X^i]|^2}+\sum_{i,j}{Tr|[X^i,X^j]|^2}}\Bigr]\Bigr)
\eea

In the limit $R_9\rightarrow 0$, to analyse the light spectrum, we must set
$E=0$ and
also set the potential terms to zero. The vanishing
of $E$ again implies that the effective theory will be described in terms of
gauge-singlet objects .

The vanishing of the potential terms implies that we are on the moduli space.
In this case,
the moduli space is more complicated; we have both a Higgs and a Coulomb
branch.
We will analyse them separately.

The Higgs branch is the branch where $\psi\neq 0$. Since $\psi$ denotes the
motion
within the ALE space, this implies that the D-branes have moved off the fixed
point.
The $Z_2$ projection implies that the images must then be at the same
point in the transverse space i.e. the D-brane and its image are at positions
$(x_2,x_3,x_4,x_5,x_6,x_7,x_8,x_9)$ and $(x_2,x_3,x_4,x_5,-x_6,-x_7,-x_8,-x_9)$
respectively.

In general, different pairs of D-branes do not have to be at the same value of
$(x_2,x_3,x_4,x_5)$. We can therefore separate the $2N$ D-branes into
$N$ pairs.

For each pair of D-branes, we need to find a set of gauge-invariant coordinates
parametrizing their position on the Higgs branch.  We must first impose
the condition that the potential energy vanishes. We must then choose a
gauge invariant set of coordinates.

It was shown in \cite{dm,jm} that after these operations, we have precisely
four gauge-invariant
scalars $Y^a$ left.
One can then find the dynamics for these scalars. It was shown in \cite{dm,jm}
that these
scalars propagate according to the Lagrangian $g_{ab}\del Y^a\del Y^b$, where
$g_{ab}$ coincides with the ALE metric: $g_{ab}=g_{ab}^{ALE}$.

This is true for a general ALE space, i.e. after imposing the vanishing of the
potential terms and gauge invariance, we have four gauge-invariant coordinates
which have a nontrivial moduli space metric, which is identical to the
ALE metric.

We now turn to the transverse coordinates. Each pair of D-branes was at
a transverse coordinate $(x_2,x_3,x_4,x_5)$. If we take $N$ pairs of 0-branes,
we
will have a $N \times N$ matrix parametrizing these positions. The gauge
invariant objects are the eigenvalues of these matrices.

The dynamics of these eigenvalues will determine the propagation of the string
in the transverse coordinates. Accordingly, one expects a trivial metric in the
kinetic terms. We have not been able to prove this directly, but this seems
very plausible for several reasons. The eigenvalue dynamics is a 1+1 sigma
model. It is well known that such a theory flows to a Ricci flat metric in the
IR. Furthermore, at the classical level, the metric is trivial. There are no
one-loop corrections \cite{ods}. Supersymmetry then implies that the flat
metric
is not corrected perturbatively. It seems reasonable to suppose that the
non-perturbative corrections will still allow the sigma model to flow to a
free theory in the IR. We shall therefore make this assumption.

As before, we have a $Z_N$ gauge symmetry. We can therefore produce long
strings by identifying the matrices upto a permutation of eigenvalues when we
go around the YM circle. These strings will have a world-sheet
Lagrangian
\be
L=(\del X^i)^2+ g_{ab}^{ALE}\del Y^a\del Y^b + fermions
\ee
which is precisely that of a string propagating in an ALE background.

We turn now to the Coulomb branch. On this branch, the $\psi^a$ are set equal
to zero, which means that the 1-branes are stuck to the fixed point. The
$X^i$ can have arbitrary values. These states were shown
\cite{dm,Polch,enhance} to
correspond to the extra gauge bosons needed to enhance
the gauge symmetry to a nonabelian group.

Hence we have obtained all the states of type IIA string theory in the
background of an ALE space.
The consistency of string theory will then automatically provide the force law
between two gravitons. Although this is a consistent theory
for finite $N$ \cite{DLCQ}, the full Lorentz invariance is
only recovered in the large $N$ limit. One should therefore expect to
recover   the force law between
two gravitons only in this limit..

One must also show that the structure of the vertex is of the
same form as in string theory. To do this requires a more detailed
analysis of the superconformal field theory. We expect that supersymmetry
is sufficient to recover the vertex structure, but a demonstration
of this is desirable.

So far we have only looked at the orbifold limit. The case where the orbifold
singularity is blown up was also considered in \cite{dm,jm}, where it was shown
that the
blowup corresponded to the addition of Fayet-Iliopoulous D-terms. Remarkably,
the result
is very similar; the metric on the Higgs branch is altered so that it is still
identical to the metric on the nonsingular ALE space! Thus everything we
have said goes through unchanged. We still recover string theory in an
ALE background, and the force law is guaranteed to work in the large $N$
limit.

When we have the full $K3$ instead of an ALE space, there are certain
complications
that arise. The compactification of M-theory on
$K3$ seems to be described by a 4+1 dimensional field theory,
like the case of compactification on $T^4$ which is a 4+1 YM theory.
These theories are difficult to define.

Presently, we can only discuss M-theory near the orbifold points of
$K3$ where  the dynamics is simple enough to enable us to make statements. At
these points,
the $K3$ is flat everywhere except near the orbifold singularities, where
the geometry is that of the ALE space. We have already seen that  0-branes
moving in an ALE space see the ALE space geometry. It is obvious that
0-branes in flat space see flat space geometry. Putting these facts
together, we see that, at least near the orbifold limit, the
D-branes see the geometry of $K3$.

We can then go through the procedure of constucting strings on $K3$ in
exact parallel to the discussion for the ALE space.The only difference is
that on the Higgs branch, the moduli space metric is the $K3$ metric instead of
the ALE metric.
Thus the strings we construct will have exactly the worldsheet action of
strings in
the background of $K3$.
This again shows that the force law between two gravitons will work out
correctly in the large $N$ limit.

\section{M-theory-heterotic duality}

In the limit where the volume of the $K3$ shrinks to zero,
we expect to recover heterotic string theory on $T^3$ \cite{witten}. We now
show
how this occurs in M(atrix) theory.

The idea is very simple. We are looking at the dynamics of \td \ in the
background of a shrinking $K3$. It is natural
to T-dualize this theory so that it becomes a theory of 4-branes wrapped on
$K3$. It turns out that these 4-branes are at strong coupling, so it is
natural to interpret this as a theory of M-theory 5-branes wrapped on
$K3$. The extended dimension of the 5-branes is much longer than the $K3$
dimensions.
We can therefore treat it as effectively being a 1+1 dimensional theory, with
a Lagrangian following from wrapping a 5-brane on $K3$.
This is known to have the same world-sheet structure as a heterotic
string \cite{HS,Cherk}. We will therefore recover the heterotic string theory
in
this limit, exactly as expected.

We would like to see that the tension of the heterotic string obtained in this
manner is
exactly the value expected from M-theory-heterotic duality.

To simplify the calculation, we first consider toroidal compactification along
the
lines of \cite{shrink},
where M-theory is compactified on a four-torus with sides $L_i,i=1,2,3,4$.
This is represented in M(atrix) theory by a 4+1 dimensional Yang-Mills
theory on a four torus of radii $\S_i=(2\pi)^2{l_{11}^3\over RL_i},i=1,2,3,4$,
with
a coupling constant $g^2=(2\pi)^6{l_{11}^6 \over RL_1L_2L_3L_4}$.

It was shown in \cite{rozali} that the coupling constant becomes an extra
dimension. Thus we reach
a 5+1 dimensional theory. This is interpreted as the theory on the
world-volume of $N$ M-theory 5-branes wrapped on a 5-torus with sides
$\S_i=(2\pi)^2{l_{11}^3\over RL_i},i=1,2,3,4$, and $\S_5=(2\pi)^5{l_{11}^6\over
RL_1L_2L_3L_4}$.

This 5+1 dimensional theory is still somewhat mysterious, but we can
still extract some information. In particular, if we shrink four out of the
five
dimensions to zero size, we reach a 1+1 dimensional theory describing
a string. We can do this in five ways, thus obtaining five strings. Four of
these
correspond to membranes wrapped on the four circles of the $T^4$. The
fifth is the 5-brane wrapped on $T^4$. There is a symmetry among these five
strings. This is related to the symmetry used in \cite{halyo,Berkroz} to give a
construction of the five-brane wrapped on $T^5$.

We already know the tensions of the four strings corresponding to
the four membranes. The string tensions corresponding to these
strings are given by $T_s={1\over R\S_i},i=1,2,3,4.$ The symmetry then
tells us that the fifth string will have tension $T_s={1\over R\S_5}={
L_1L_2L_3L_4\over (2\pi)^5l_{11}^6}$,
which is the right tension for a five-brane wrapped on $T^4$.

We can perform exactly the same manipulations for M-theory on a $K3$ of
volume $vl_{11}^4$. This, according to our conjecture, is a theory
of \td \ moving on this $K3$. After T-duality\cite{oz}, this is described by a
4+1
dimensional theory on a $K3$ of volume ${l_{11}^{12}\over
R^4(vl_{11}^4)}={l_{11}^8\over
R^4v}$. This is equivalent to a 5+1 dimensional theory, which we interpret
as the world-volume of $N$ M-theory 5-branes wrapped on $K3\times S^1$. The
length of the $S^1$ is, as before, $\S_5=(2\pi)^5{l_{11}^2\over Rv}$.
In the limit $v\rightarrow 0$, this is a 1+1 dimensional theory where the
5-branes
are wrapped on a small $K3$ and a large $S^1$.

Although we do not know the world-volume theory of the 5-brane, it is
very reasonable to assume that the dynamics will force us onto the moduli
space of the theory. This means that the $N$ 5-branes will be separated
from each other in the transverse space. Each 5-brane is wrapped on a
very small $K3$, leading to a 1+1 dimensional theory which
is known to be a theory
which describes the worldsheet action of the heterotic string
\cite{HS,Cherk}.
As usual we have a $Z_N$ symmetry, and we can go through the
usual procedure of making long strings.

The string scale of
the heterotic string is, as before, $T_s={1\over R\S_5}={ v\over
(2\pi)^5l_{11}^2}$,
which is the correct value predicted from string duality.

It would be interesting to see the connection of this construction to the
matrix model
proposed for  heterotic strings \cite{banksmotl}.

\section{Acknowledgements}

We are grateful to E.Halyo and especially to L.Susskind for crucial
discussions.

W. F. thanks the ITP at Stanford for its hospitality and was
supported in part by the Robert Welch Foundation and
NSF Grant PHY-9511632. A.R. was supported in part by
the Department of Energy
under contract no. DE-AC03-76SF00515.




\end{document}